\documentclass[%
 reprint,
 amsmath,amssymb,
 aps,
]{revtex4-1}

\usepackage{xcolor}
\usepackage{graphicx}
\usepackage{dcolumn}
\usepackage{bm}

\begin{document}

\title{Swimmers' wake structures are not reliable indicators of swimming performance}

\author{Daniel Floryan}
\email{dfloryan@alumni.princeton.edu}
\author{Tyler Van Buren}
\author{Alexander J. Smits}
\affiliation{
  Department of Mechanical and Aerospace Engineering, Princeton University, Princeton, NJ 08544, USA
}%

\date{\today}

\begin{abstract}
The structure of swimmers' wakes is often assumed to be an indicator of swimming performance, that is, how momentum is produced and energy is consumed. Here, we discuss three cases where this assumption fails. In general, great care should be taken in deriving any conclusions about swimming performance from the wake flow pattern. 
\end{abstract}

\maketitle

\section{Introduction}
\label{sec:intro}

When swimmers propel themselves through a fluid, they leave a distinct pattern of fluid flow in their wakes analogous to the footprints of terrestrial animals \citep{zhang2017footprints}. The flow structures in the wake carry much information with them; for example, a complete control volume analysis can translate the velocity and stress fields in the fluid to forces on the swimmer, although the number of detailed measurements required usually makes this analysis impractical for any experimental investigation even for two-dimensional flows (see Appendix~\ref{sec:app}).  Through the advent of particle image velocimetry in particular, flow structures are furthermore much more accessible to scientists than the musculature of swimmers, offering a non-intrusive way to estimate the forces produced and energy expended by swimmers. This offers some motivation for why, it seems, studies of swimming animals so often show the flow structures in the swimmers' wakes \citep{muller1997fish}. (They are also quite visually-pleasing.) 

The present letter, however, is a cautionary one. Here, we re-interpret results in the literature and offer three cases demonstrating that the wakes of swimmers can be entirely misleading when trying to assess the propulsive performance of swimmers. We give focus to fast and efficient swimmers, epitomized by animals such as tuna that have characteristically large-aspect-ratio tails that generate nearly all of the propulsive force and are well-separated from the rest of the body (which essentially amounts to a source of balancing drag) \citep{webb1984form, smits2019undulatory}. The essential features of the propulsion of these swimmers can therefore be modeled by a flapping rectangular foil \citep{wu2011fish}, which represents the isolated propulsive surface of these animals and is henceforth referred to as the propulsor. We consider propulsors moving with a constant free-stream velocity; for the swimmers of interest, constant velocity studies can be used to make robust conclusions about free (untethered) swimming, where the swimmers are allowed to accelerate \citep{van2018flow}. 

In the first case, we show that significant changes in the wake can be associated with no changes in propulsion (specifically thrust production); in the second case, we show that small changes in the wake can be associated with large changes in propulsion (specifically peak efficiency); and in the third case, we show that changes in the pattern and self-interaction of the wake are associated with changes in propulsion (thrust production, power consumption, and efficiency) that are captured by simple models that are agnostic to the state of the wake.

\section{First case: vortex spacing}
\label{sec:first}

\begin{figure}
  \begin{center}
  \includegraphics[width=0.7\linewidth]{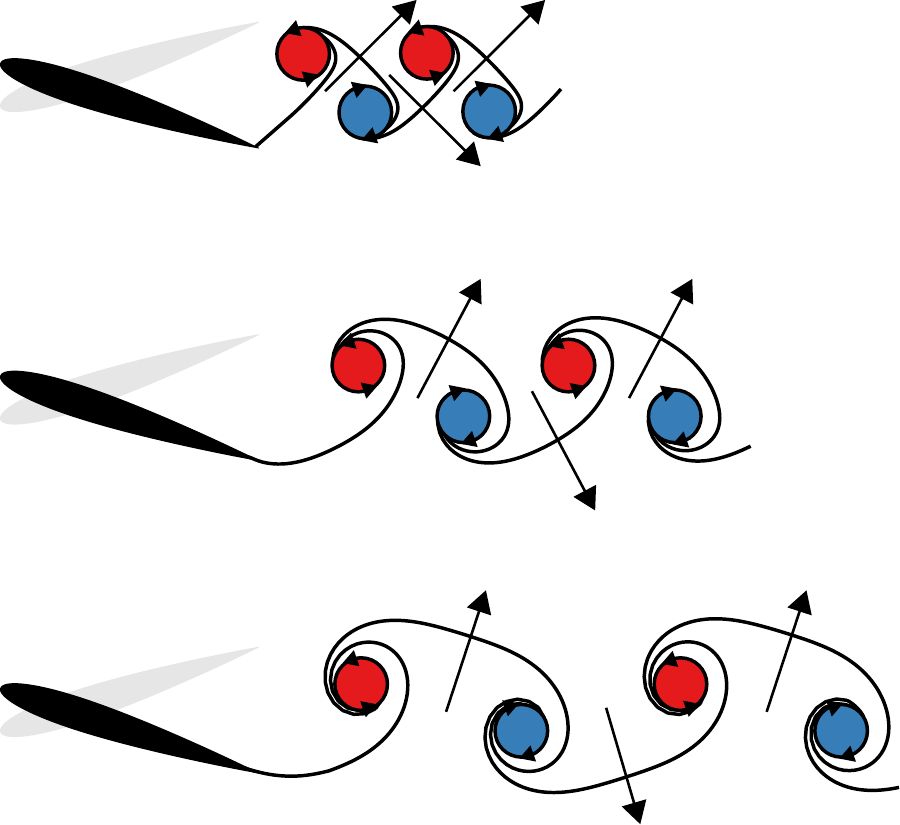}
  \end{center}
  \caption{Thrust-producing wakes with increasing horizontal spacing between vortices. }
  \label{fig:spacing}
\end{figure}

An often-cited mechanism for how flapping propulsors produce thrust is based on the wake mechanics \citep{triantafyllou1991wake}.  As a propulsor flaps and moves forward, it leaves behind a staggered array of vortices in its wake.  A common pattern is shown schematically in figure~\ref{fig:spacing}, where two opposite-sign vortices are shed per flapping period, arranged such that those with a counterclockwise orientation are positioned above those with a clockwise orientation (other patterns are possible; we will return to this point later). The vortices induce flow that takes the form of a meandering jet that increases the streamwise momentum of the fluid opposite to the direction of travel. By action-reaction, the fluid imparts a thrust force onto the propulsor. 

Based on this mechanism, it is intuitive that the spacing of the vortices should dictate how much thrust is produced since the spacing dictates the intensity and direction of the induced flow. Suppose we run three experiments: in each experiment, the flapping frequency and amplitude are the same, but from one experiment to the next we increase the forward speed of the propulsor.  What happens to the thrust? The resulting wakes are sketched in figure~\ref{fig:spacing}, with speed increasing from top to bottom (note that the strengths and sizes of the vortices do not change under these circumstances \citep{buchholz2011scaling, green2011unsteady}). When the vortices are spaced closer horizontally, the induced velocity has a larger component in the streamwise direction, and we should expect that more thrust is produced. 

In fact, despite significant changes in the spacing of the vortices in the wake, all three experiments will produce the same mean thrust (although the efficiencies will be different; see \citet{floryan2018efficient} for details).  It has been shown indirectly \citep{gazzola2014scaling} and directly \citep{van2018flow, floryan2018efficient} that for flapping motions representative of those seen in nature, the mean thrust $T$ is independent of the swimming speed of the animal, instead being proportional to the density of the fluid $\rho$, the area of the propulsor $S$, and the square of the velocity of the trailing edge $V$, so that 
\begin{equation}
  \label{eq:thrust}
  T \sim \rho V^2 S,
\end{equation}
where $\sim$ indicates a proportionality. Here, we distinguish the thrust $T$ from the net force $F_x = T - D$, where $D$ is an offset drag \citep{gazzola2014scaling, floryan2018efficient}. In the three experiments we have described, the horizontal spacing of the vortices is dictated by the ratio of the swimming speed to the frequency of flapping, $U_\infty/f$. Since the mean thrust is independent of the swimming speed, it should be clear that the spacing of vortices in the wake cannot reliably give an indication of the mean thrust produced. 

A subtle point needs to be addressed. If we non-dimensionalize thrust by the dynamic pressure and area of the propulsor, $\rho U_\infty^2 S$, as is often done for forces due to fluids, then the normalized thrust scales as $(V/U_\infty)^2$, the square of a velocity ratio. Rewriting the trailing edge velocity $V$ as the product of flapping amplitude $A$ and flapping frequency, the dimensionless thrust scales as $(A/(U_\infty/f))^2$, the square of a spacing ratio. In fact, this spacing ratio gives approximately the ratio of the vertical spacing of the vortices in the wake to their horizontal spacing. The dimensionless thrust therefore depends directly on the spacing of vortices in the wake, whereas the dimensional thrust does not. Has something gone awry? 

In fact, we are dealing with a tautology. If we take a quantity that is independent of $U_\infty$---such as the dimensional thrust $T$---and divide it by another quantity that depends on $U_\infty$, then the result will depend on $U_\infty$ by construction. This merely points to the fact that the usual dynamic pressure is not the appropriate quantity by which to non-dimensionalize forces generated by swimmers. Non-dimensionalizing by $\rho V^2 S$ instead gives that the dimensionless thrust is constant and therefore independent of the vortex spacing, in concert with the dimensional thrust. It should be clear that the spacing of vortices in the wake cannot reliably give an indication of the mean thrust produced. 

\begin{figure}
  \begin{center}
  \includegraphics[width=0.55\linewidth]{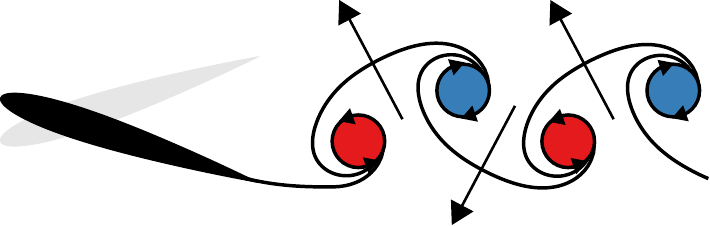}
  \end{center}
  \caption{Drag-producing wake. }
  \label{fig:drag}
\end{figure}

The vertical spacing of vortices in the wake is also commonly used to make conclusions about thrust production.  As described previously, when the counterclockwise-oriented vortices are positioned above the clockwise-oriented vortices, we expect the propulsor to produce thrust; a wake with this arrangement of vortices is called a reverse von K\'arm\'an vortex street and is often termed a ``thrust-type'' wake \citep{sfakiotakis1999review}. Following the same logic, we expect a propulsor to produce drag when the vortices have the opposite arrangement, as sketched in figure~\ref{fig:drag}; a wake with this arrangement of vortices is called a von K\'arm\'an vortex street and is often termed a ``drag-producing'' wake \citep{sfakiotakis1999review}. When the vortices are in line, we expect no net horizontal force; this arrangement of vortices marks the drag-thrust transition.  Following this line of argument, the vertical arrangement of vortices in the wake is often used to determine whether thrust or drag is being produced \citep{jones1998experimental}. 

Conclusions regarding thrust based on the vertical spacing of vortices are inaccurate, however. Generally speaking, the drag-thrust transition occurs when the wake is already a ``thrust-type'' wake \citep{andersen2017wake}, since some excess streamwise fluid momentum is needed to overcome profile drag or velocity fluctuations and pressure differences in the control volume \citep{ramamurti2001simulation, bohl2009mtv}. Furthermore, even ``drag-producing'' wakes have been observed to produce thrust \citep{andersen2017wake}.  In this context, the recent work by \citet{lagopoulos2019universal} offers an alternative method to distinguish drag- and thrust-producing behavior based on kinematic inputs instead of the vortex arrangement.  Despite intuition, vortex spacing, either in the horizontal or vertical direction, is not a reliable indicator of thrust production.

\section{Second case: Reynolds number}
\label{sec:second}

The Reynolds number $Re=U_\infty c/\nu$ measures the strength of inertial forces relative to the strength of frictional forces in a flow.  Here, $c$ is the chord length of the propulsor, and $\nu$ is the kinematic viscosity of the fluid.  For flapping propulsors, the effects of the Reynolds number are typically not considered (for example, in \citet{anderson1998oscillating} the authors use results from flow visualizations captured at $Re = 1\,100$ to make conclusions about swimming performance at $Re = 40\,000$). This is likely because Reynolds number effects are presumed to be small compared to kinematic effects, which are, of course, strong.  In addition, the structure of the wake typically has a rather weak dependence on the Reynolds number: in studies spanning a large range of Reynolds numbers, geometries, and kinematics, the authors have shown that the basic process of vortex formation and the establishment of the wake are not significantly affected by the Reynolds number \citep{ohmi1990vortex, ohmi1991further, dong2005wake, jantzen2014vortex, senturk2018numerical}. Increasing the Reynolds number tends to lead to the appearance of some small-scale structures and a sharpening of flow structures, but the basic sketch drawn in figure~\ref{fig:spacing} does not change. Based solely on the wake, we would not expect much of a change in swimming performance with Reynolds number. 

Nevertheless, the efficiency of propulsion turns out to be quite sensitive to the Reynolds number, especially with regard to its optimal value. (Here, we use the Froude efficiency $\eta = F_x U_\infty/P$, where $P$ is power consumption.) The efficiency's sensitivity to the Reynolds number was shown analytically in \citet{floryan2018efficient} and confirmed by simulations in \citet{senturk2019reynolds}. To explain why this is so, we first note that the optimal efficiency coincides with low net thrust (see \citet{floryan2018efficient} for details). Changing the Reynolds number will lead to a small change in net thrust, which changes the efficiency by
\begin{equation}
  \label{eq:eta}
  \Delta \eta = \frac{\partial \eta}{\partial F_x} \Delta F_x = \frac{\eta}{F_x} \Delta F_x.
\end{equation}
Even though the change in net thrust may be small, the change in optimal efficiency will be large because it coincides with low net thrust.  Changing the Reynolds number may not change the wake much, which would lead us to believe that swimming performance is hardly affected, but the efficiency may change substantially. The wake is therefore not a reliable indicator of efficiency.

\section{Third case: vortex pattern and interactions}
\label{sec:third}

\begin{figure}
  \begin{center}
  \includegraphics[width=0.7\linewidth]{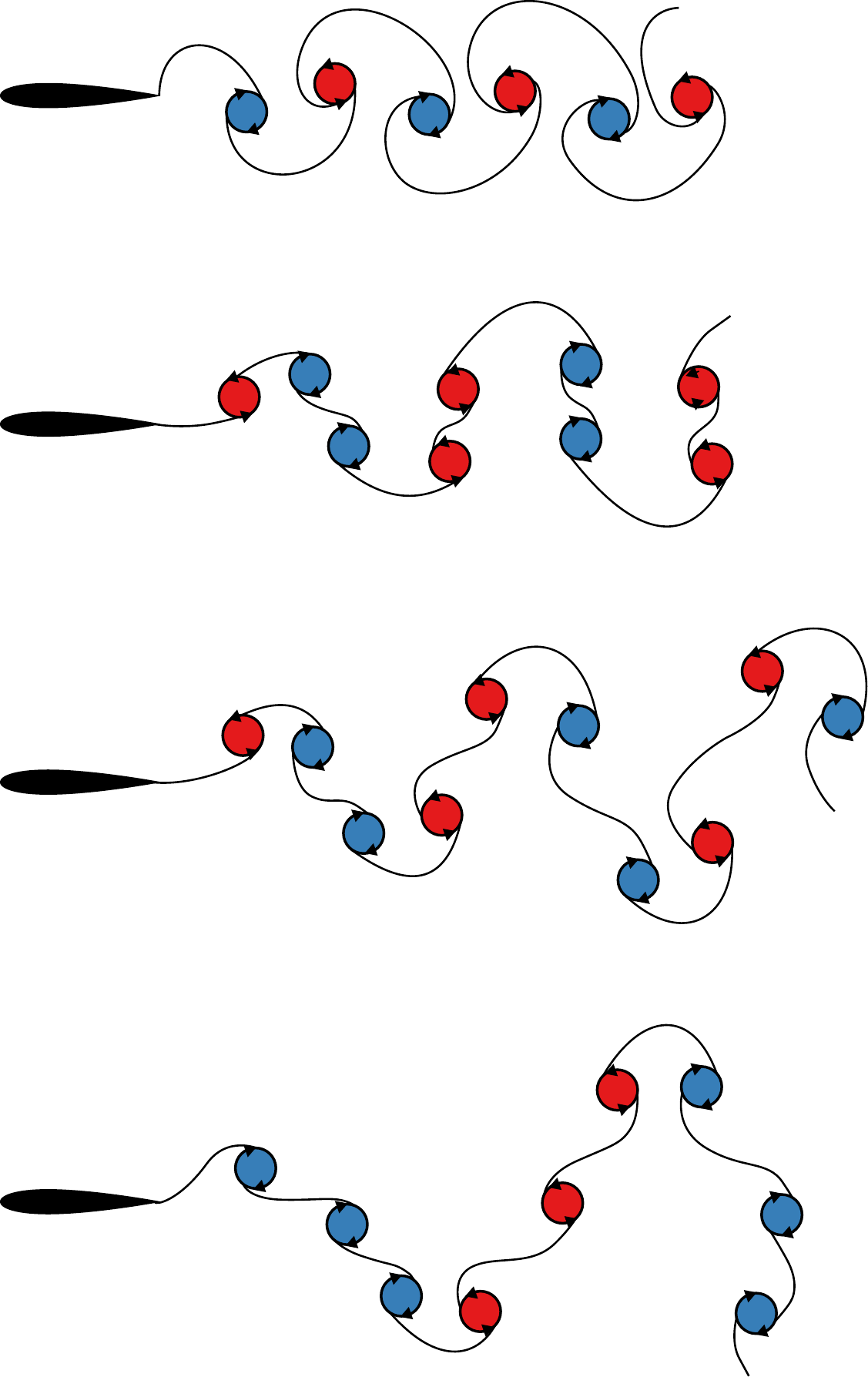}
  \end{center}
  \caption{Vortex patterns from \citet{andersen2017wake}. }
  \label{fig:patterns}
\end{figure}

Although the reverse von K\'arm\'an vortex street is the most commonly encountered wake vortex pattern, many other patterns are possible. In figure~\ref{fig:patterns} we have sketched some of the patterns observed in the experiments and computations of \citet{andersen2017wake}; even more exotic patterns have been observed. With wildly varying wake patterns, we may expect to see large differences in swimming performance as the wake transitions from one pattern to another. 

In fact, we do not. There is ``no evidence of particular vortex patterns having a distinct effect on force measurements'' \citep{mackowski2015direct}. As kinematic parameters are varied, all metrics of swimming performance vary smoothly, even when the wake transitions from one pattern to another. The insensitivity of swimming performance to the type of vortex pattern observed is further supported by the ability of simple models---models that are agnostic to the vortex pattern---to capture swimming performance as a function of the kinematics \citep{floryan2017scaling, van2018scaling, floryan2018efficient}. 

The vortex pattern arises as a consequence of the self-interaction of the vortices in the wake. This self-interaction does not support a net force; rather, the force on the propulsor is due to the vortical impulse associated with the shedding of vorticity from the propulsor \citep{wu1981theory}. The creation of vorticity, not its subsequent evolution, is important, explaining the insensitivity of swimming performance to vortex pattern. For instance, \citet{young2007mechanisms} compared the swimming performance of flapping propulsors whose wakes were allowed to deform according to the induced velocity field with those that were not. The authors found no difference in swimming performance, indicating that although the location where vorticity is shed (the trailing edge) is important, the subsequent development of the vortex pattern is not. This suggests that the interactions between vortices have little bearing on swimming performance. 

The importance of vortex interactions was also addressed in \citet{floryan2017forces}. There, the authors investigated intermittent swimming motions, where the propulsor alternates between one period of flapping and rest; the duty cycle gives the proportion of time spent flapping.  Each burst of flapping releases a group of vortices into the wake, and the duty cycle determines the spacing between the groups. When the duty cycle is low, groups of vortices are independent of each other. As the duty cycle increases, the groups of vortices move closer and should interact more strongly; in this sense, the duty cycle provides a way to control the strength of vortex interactions. The authors found that the time-averaged  thrust and power (averaged over flapping and rest time) simply scaled linearly with the duty cycle, that is, they are independent of duty cycle when averaged only over  the time that the propulsor flaps. The individual bursts of vortices are therefore effectively independent of each other even as the duty cycle tends toward unity (at least as far as swimming performance is concerned). Vortex interactions, and the resulting vortex patterns, apparently have little bearing on swimming performance.

\section{Conclusion}
\label{sec:conc}

The wake behind swimmers is often looked to as an indicator of swimming performance. We have offered three cases to the contrary. Significant differences in the wake may cause no changes in swimming performance, insignificant changes to the wake may cause great changes in swimming performance, and the pattern and self-interaction of the wake have little bearing on swimming performance, dispelling the notion that there is a preferred pattern of vortices. Admittedly, we are also guilty of looking to the wake to explain swimming performance \citep{van2017impact}. This is not to say that the wake is not informative (indeed, swimming performance can be recovered from wake measurements when a control volume analysis is properly performed, but doing so requires information that is difficult to obtain in experiments, e.g., velocity-pressure correlations); we merely point out that conclusions based on the wake can be misleading, and that great care should be taken. 

There are other cases when the wake is critical to understanding the problem, but these involve the wake of one swimmer impinging on another swimmer (or, more generally, a swimmer negotiating an unsteady incoming flow). For example, \citet{wu1972extraction} provides the theoretical basis to show that an unsteady incoming flow can lead to a scenario where a flapping propulsor can simultaneously produce thrust and extract energy on average, which is not possible with a steady uniform incoming flow, and \citet{beal2006passive} showed in experiments that a flapping propulsor can produce thrust and extract energy on average in the presence of an incoming vortex wake. It is also apparent that live trout take advantage of incoming vortices to expend less energy \citep{liao2003fish}. Indeed, the upstream flow may be quite important for a swimmer, but the downstream flow should be treated with great care as its analysis may mislead us. \\

This work was supported by ONR Grant N00014-14-1-0533 (Program Manager R. Brizzolara).

\appendix

\section{Control volume analysis}
\label{sec:app}

The force production and energy consumption of flapping foils can be extracted from the velocity and stress fields. Applying conservation of momentum to a control volume gives
\begin{equation}
  \label{eq:mom}
  \frac{\partial}{\partial t} \int_V \rho \mathbf{u}\,\text{d}V + \int_S (\hat{\mathbf{n}} \cdot \rho \mathbf{u}) \mathbf{u}\,\text{d}S = \int_S \mathbf{\Sigma}\,\text{d}S + \mathbf{R}_{\text{ext}},
\end{equation}
where $V$ is the volume, $S$ is the surface, $\hat{\mathbf{n}}$ is the outward-facing unit normal on the surface, $\cdot$ denotes the dot product, $\rho$ is the density of the fluid, $\mathbf{u}$ is the velocity field of the fluid, $\mathbf{\Sigma}$ is the traction vector, and $\mathbf{R}_\text{ext}$ is the external force acting on the fluid in the control volume. The traction vector is given by
\begin{equation}
  \label{eq:trac}
  \Sigma_j = \sigma_{ij} n_i,
\end{equation}
where $\sigma_{ij}$ is the stress tensor and the Einstein summation convention is used. We assume incompressible flow, so that 
\begin{equation}
  \label{eq:stress}
  \sigma_{ij} = -p \delta_{ij} + \mu \left( \frac{\partial u_i}{\partial x_j} + \frac{\partial u_j}{\partial x_i} \right),
\end{equation}
where $p$ is the pressure and $\delta_{ij}$ is the Kronecker delta. 

\begin{figure}
  \begin{center}
  \includegraphics[width=0.7\linewidth]{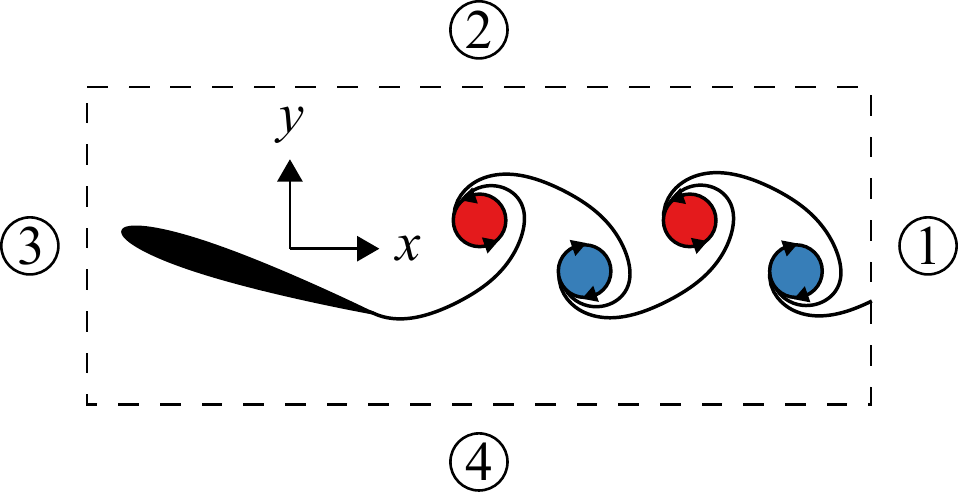}
  \end{center}
  \caption{Control volume. }
  \label{fig:cv}
\end{figure}

For convenience, we decompose the flow variables into their time-averaged and fluctuating components, with averaging performed over an integer number of periods of flapping. The time-averaged component is denoted by an overbar, and the fluctuating component is denoted by a prime. Applying~\eqref{eq:mom} for two-dimensional flow to the control volume sketched in figure~\ref{fig:cv}, taking its streamwise component, and time-averaging gives
\begin{eqnarray}
  \label{eq:momcv}
  &&\int_1 \rho \overline{u}^2\,\text{d}S + \int_1 \rho \overline{u'^2}\,\text{d}S + \int_2 \rho \overline{u}\,\overline{v}\,\text{d}S+ \int_2 \rho \overline{u' v'}\,\text{d}S \nonumber \\
  &&- \int_3 \rho \overline{u}^2\,\text{d}S - \int_3 \rho \overline{u'^2}\,\text{d}S - \int_4 \rho \overline{u}\,\overline{v}\,\text{d}S - \int_4 \rho \overline{u' v'}\,\text{d}S \nonumber \\
  =&& \int_1 \left( -\overline{p} + 2\mu\frac{\partial \overline{u}}{\partial x} \right)\,\text{d}S + \int_2 \mu \left( \frac{\partial \overline{v}}{\partial x} + \frac{\partial \overline{u}}{\partial y} \right)\,\text{d}S \nonumber \\
  &&- \int_3 \left( -\overline{p} + 2\mu\frac{\partial \overline{u}}{\partial x} \right)\,\text{d}S - \int_4 \mu \left( \frac{\partial \overline{v}}{\partial x} + \frac{\partial \overline{u}}{\partial y} \right)\,\text{d}S \nonumber \\[1mm]
  &&+ \overline{F_x},
\end{eqnarray}
where $F_x$  is the external streamwise force acting on the fluid and the limits of integration refer to the numbered sides in figure~\ref{fig:cv}. If the velocity field, its gradient, and the pressure are known accurately on the surface of the control volume, the mean thrust can be recovered. 

Applying conservation of energy to a control volume gives
\begin{equation}
  \label{eq:en}
  \frac{\partial}{\partial t} \int_V \frac{1}{2} \rho \mathbf{u} \cdot \mathbf{u}\,\text{d}V + \int_S \frac{1}{2}\left( \hat{\mathbf{n}} \cdot \rho \mathbf{u} \right) \mathbf{u} \cdot \mathbf{u}\,\text{d}S = \int_S \mathbf{\Sigma}\cdot \mathbf{u}\,\text{d}S+ \dot{W},
\end{equation}
where $\dot W$ is the rate of work done on the system. 
For the control volume shown in figure~\ref{fig:cv} we obtain
\begin{eqnarray}
  \label{eq:encv}
  &&\frac{1}{2}\int_1 \rho \overline{u(u^2 + v^2)}\,\text{d}S + \frac{1}{2}\int_2 \rho \overline{v(u^2 + v^2)}\,\text{d}S \nonumber \\
  &&- \frac{1}{2}\int_3 \rho \overline{u(u^2 + v^2)}\,\text{d}S - \frac{1}{2}\int_4 \rho \overline{v(u^2 + v^2)}\,\text{d}S \nonumber \\
  =&& \int_1 \left[ -\overline{pu} + 2\mu\overline{\frac{\partial u}{\partial x} u} + \mu \overline{\left( \frac{\partial u}{\partial y} + \frac{\partial v}{\partial x} \right) v}\right]\,\text{d}S  \nonumber \\
  &&+ \int_2 \left[ -\overline{pv} + 2\mu\overline{\frac{\partial v}{\partial y} v} + \mu \overline{\left( \frac{\partial u}{\partial y} + \frac{\partial v}{\partial x} \right) u}\right]\,\text{d}S  \nonumber \\
  &&- \int_3 \left[ -\overline{pu} + 2\mu\overline{\frac{\partial u}{\partial x} u} + \mu \overline{\left( \frac{\partial u}{\partial y} + \frac{\partial v}{\partial x} \right) u}\right]\,\text{d}S  \nonumber \\
  &&- \int_4 \left[ -\overline{pv} + 2\mu\overline{\frac{\partial v}{\partial y} v} + \mu \overline{\left( \frac{\partial u}{\partial y} + \frac{\partial v}{\partial x} \right) u}\right]\,\text{d}S  \nonumber \\[1mm]
  &&+ \overline{\dot W},
\end{eqnarray}
where terms expand as
\begin{eqnarray}
  \label{eq:exp}
  \overline{pu} & =& \overline{p}\,\overline{u} + \overline{p' u'}, \\
  \overline{u \frac{\partial u}{\partial x}} &=& \overline{u}\frac{\partial \overline{u}}{\partial x} + \overline{u' \frac{\partial u'}{\partial x}}, \\
  \overline{u(u^2 + v^2)} & = & \overline{u}^3 + 3\overline{u}\overline{u'^2} + \overline{u}\,\overline{v}^2 + \overline{u}\overline{v'^2} \nonumber\\
  &&+ \overline{u'^3} + 2\overline{v}\overline{u' v'} + \overline{u' v'^2},
\end{eqnarray}
and similarly for other terms.  Full knowledge of the velocity and stress fields on the surface of the control volume therefore furnishes the mean power consumption. 

It is clear that using a control volume approach to obtain force production and power consumption experimentally is a great challenge. Each quantity must be known precisely, which is particularly difficult for the derivative terms.  In addition, accurately estimating the pressure field from velocity data is an area of study in its own right \citep{van2013piv}, and fraught with uncertainty. We note that some recent works have made efforts to circumvent the difficulties in using a full control volume analysis, developing methods to improve the estimation of forces from the wake \citep{dabiri2005estimation, wang2014evaluation, wang2019estimating}.

\bibliography{references}

\end{document}